\documentclass[%
 reprint,
superscriptaddress,
 amsmath,amssymb,
 aps,
]{revtex4-2}

\usepackage{gensymb}
\usepackage{ulem}
\usepackage{xcolor}
\usepackage{graphicx}
\usepackage{dcolumn}
\usepackage{bm}
\usepackage[hidelinks]{hyperref}

\begin{document}

\preprint{APS/123-QED}

\title{Deterministic Laser Writing of Spin Defects in Nanophotonic Cavities}

\author{Aaron M. Day}
    \thanks{Equal Contribution}
    \affiliation{John A. Paulson School of Engineering and Applied Sciences, Harvard University, Cambridge, Massachusetts 02138, USA}
\author{Jonathan R. Dietz}
    \thanks{Equal Contribution}
    \affiliation{John A. Paulson School of Engineering and Applied Sciences, Harvard University, Cambridge, Massachusetts 02138, USA}
\author{Madison Sutula}
    \affiliation{Department of Physics, Harvard University, Cambridge, Massachusetts 02138, USA}
\author{Matthew Yeh}
    \affiliation{John A. Paulson School of Engineering and Applied Sciences, Harvard University, Cambridge, Massachusetts 02138, USA}
\author{Evelyn L. Hu}
    \affiliation{John A. Paulson School of Engineering and Applied Sciences, Harvard University, Cambridge, Massachusetts 02138, USA}

\date{\today}

\begin{abstract}
High-yield engineering and characterization of cavity-emitter coupling is an outstanding challenge in developing scalable quantum network nodes. Ex-situ defect formation processes prevent real-time defect-cavity characterization, and previous in-situ methods require further processing to improve emitter properties or are limited to bulk substrates. We demonstrate direct laser-writing of cavity-integrated spin defects using a nanosecond-pulsed above-bandgap laser. Photonic crystal cavities in 4H-silicon carbide serve as a nanoscope monitoring silicon monovacancy (V$_{Si}^-$) defect formation within the $100~\text{nm}^3$ cavity mode volume. We observe defect spin resonance, cavity-integrated photoluminescence and excited-state lifetimes consistent with conventional defect formation methods, without need for post-irradiation thermal annealing. We further find an exponential reduction in excited-state lifetime at fluences approaching the cavity amorphization threshold, and show single-shot local annealing of the intrinsic background defects at the V$_{Si}^-$ formation sites. This real-time in-situ method of localized defect formation, paired with demonstration of cavity-integrated defect spins, marks an important step in engineering cavity-emitter coupling for quantum networking.
\end{abstract}

\maketitle

\section{Main}
Optically-active solid-state spin defects are a favorable candidate for quantum networking due to their spin-photon interface, long spin coherence, coupling to nuclear spin memories, and natural integration into photonic crystal cavities (PCC) \cite{Stas2022, Bhaskar2020, Babin2022, Lukin2020a}. PCCs augment the optical properties of defects via the Purcell effect by enhancing the emission of zero-phonon line (ZPL) photons \cite{Bracher2017,Crook2020,Gadalla2021}. Nanophotonic cavities additionally enable robust spin-photon entanglement interfaces which form the basis of quantum electrodynamic (QED) protocols \cite{Knall2022,Sipahigil2016,Lukin2022,Patton}. Defect placement is central to engineering large cavity-emitter cooperativity for optimal implementation of these proposals. Conventionally, focused ion beam implantation and masked ion implantation have been used to locally incorporate solid state spin defects \cite{FIB2022,schroder2017scalable}. Alternatively, direct laser writing using a below-bandgap femtosecond pulsed laser has recently been explored with nitrogen vacancy centers in diamond \cite{Chen2017}, V$_{Si}^-$ in 4H-silicon carbide (SiC) \cite{Chen2019}, and neutral divacancy (VV$^0$) in 4H-SiC \cite{Almutairi2022}. These previous laser-writing methods demonstrate control over defect number and radial distribution, but degradation of optical and spin properties near surfaces has limited nanophotonic integration \cite{Almutairi2018, Chen2019}. Moreover, these demonstrations have been limited to bulk materials. 

In this work, we directly write optically active spin defects into the mode volume of nanophotonic cavities using a new method of above-bandgap laser irradiation with a single nanosecond pulse. We probe the defect formation process with the nanophotonic cavity modes in real time by incorporating the irradiation laser into the confocal system, without needing to post-process or thermally anneal the formed emitters. We verify that the optical and spin properties associated with the defects are preserved by characterizing their photoluminescence and optically detected magnetic resonance (ODMR). Additionally, we gain insight on the relative damage imparted to the crystal lattice during laser irradiation by studying the relationship of fluence to emitter lifetime. Surprisingly, we also observe single-shot laser annealing, measuring a reduction in intrinsic background defects from laser irradiation.

The substrates used in these experiments are nano-scale photonic crystal cavities fabricated in unimplanted 4H-SiC using a process outlined in \cite{Bracher2017}. 4H-SiC is an attractive quantum networking material platform: it is a wide-bandgap, fabrication-amenable semiconductor commercially available at wafer scale which hosts two promising near-infrared spin defects--the VV$^0$ and V$_{Si}^-$. Additionally, 4H-SiC exhibits low acoustic loss enabling the design of devices for coherent acoustic spin control in hybrid quantum systems \cite{Dietz2022, Whiteley2019}. Recent results showing single-shot spin readout, preserved coherence of single emitters in nanophotonics, and quantum and nonlinear optics on insulator demonstrate rapid progress toward a competitive commercial quantum networking platform  \cite{Anderson2022,Babin2022,Lukin2020,Lukin2020a}. The cavities used in this work are 200~nm thick with mode volumes of approximately $100~\text{nm}^3$; as such, photoluminescence measurements in photonic structures reveal coupling to both emissive intrinsic surface defects and defects generated via laser irradiation. Unidentified surface defects postulated to be related to surface oxide are routinely observed in 4H-SiC with emission spanning the 600-1000~nm range \cite{Dietz2022a,Widmann2018,Babin2022}. 

\begin{figure}[ht!]
\includegraphics[scale = 0.93]{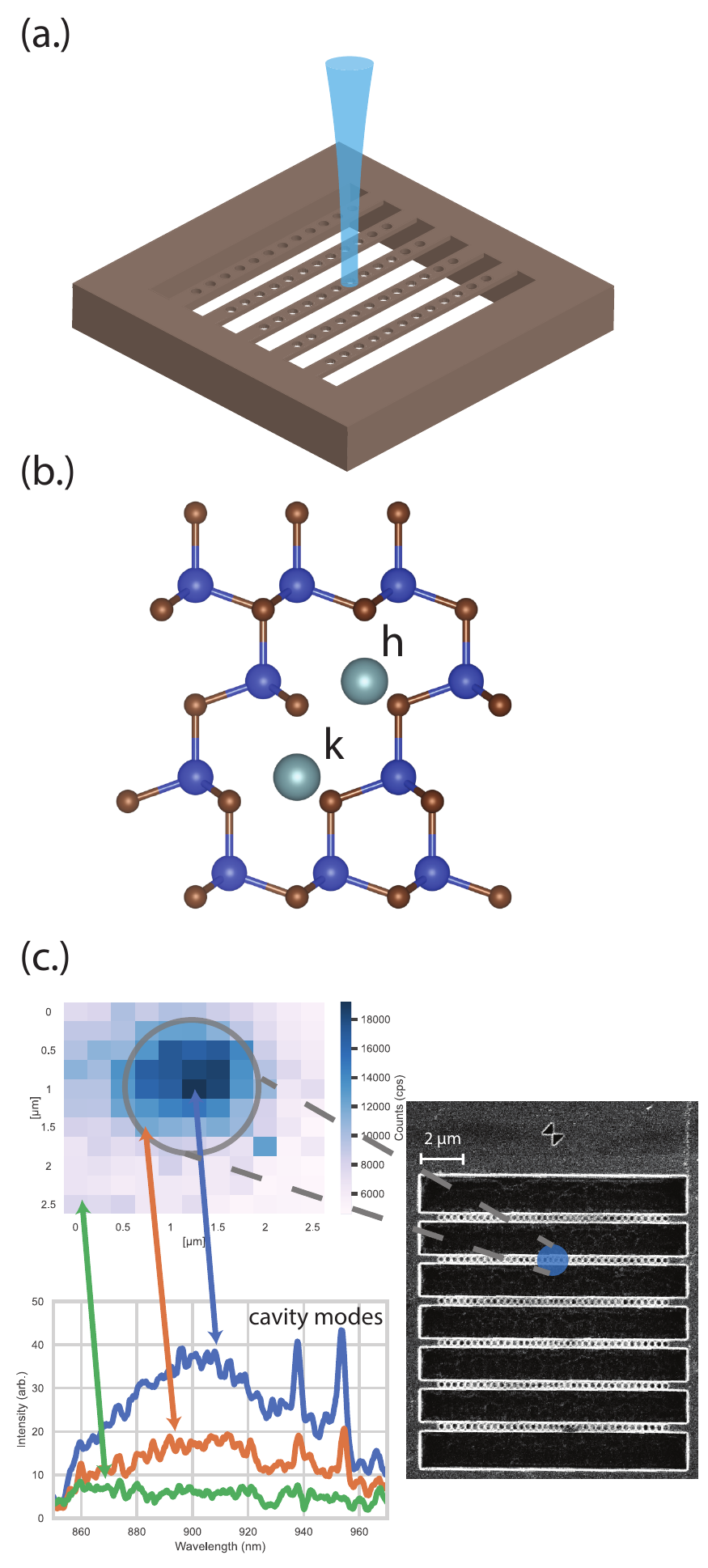}
\caption{\textbf{Laser-Writing Cavity-Integrated Spin Defects} (a) Single pulses of a 337.1~nm 4~ns pulsed laser with $\mu$J-order fluence per pulse are focused onto nanophotonic crystal cavities in 4H-SiC in order to locally incorporate (b) h and k-site silicon monovacancy defects. The silicon monovacancy is a near-IR optically-active spin-3/2 defect comprised of a missing silicon atom at an h or k lattice site surrounded by four neighboring carbon atoms. (c.) Room-temperature emission spectrum is decorated with cavity modes following localized incorporation of defects into previously uniraddiated material with intensity localized at the cavity mode site.}
\label{SystemOverview}
\end{figure}

Individual pulses of a UV (337.1~nm) laser of order nanosecond duration are focused onto the cavities (Fig.~\ref{SystemOverview}a), forming silicon monovacancy defects at the irradiation site (Fig.~\ref{SystemOverview}b). Both the individual pulse fluence and the total number of pulses delivered to the cavities are varied to determine optimal irradiation conditions which deliver laser energy greater than a minimum threshold required to observe a change in the optical signature but lower than energies that begin to amorphize the cavity \cite{Day}. For single pulse irradiation, that optimum range spanned 0.4-0.9$~\mu$J-per-pulse.  Within a $\mu$J-order range of pulse fluences and few pulse repeats,  V$_{Si}^-$ emission is observed upon irradiation, localized to the cavity mode volume and decorated by cavity modes with quality factors ranging from 2000-5500. A spatial map of photoluminescence (PL) intensity generated at room temperature is shown in Fig.~\ref{SystemOverview}c.

\begin{figure*}
\includegraphics[scale=1]{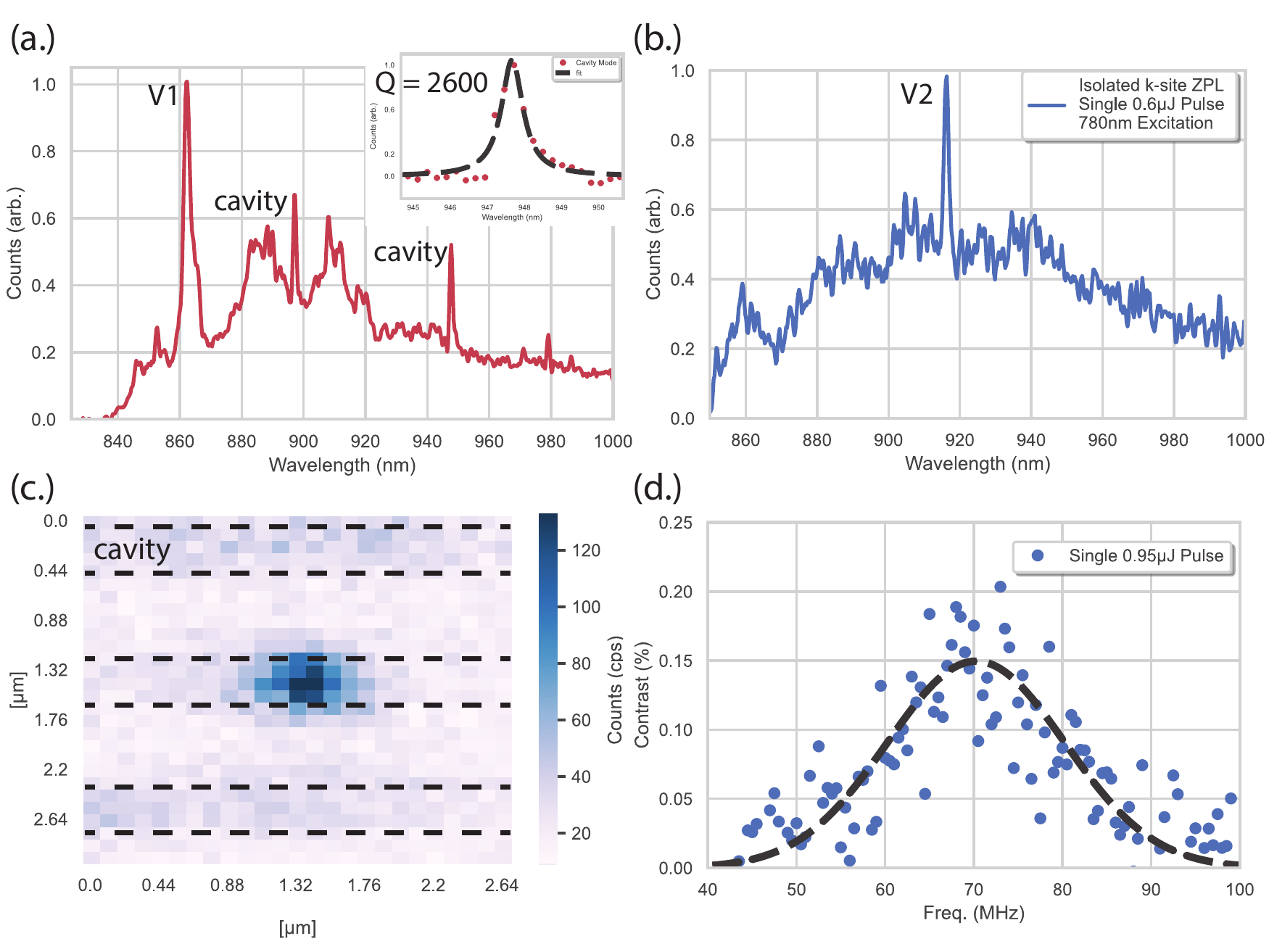}
\caption{\label{fig:wide} \textbf{Preservation of cavity mode, optical and spin signatures upon irradiation with a single UV pulse} (a.) 77 Kelvin photoluminescence spectrum of cavity-integrated h-site silicon monovacancy formed with a single 0.475~$\mu$J UV pulse, showing no observable k-site emission. The V1 zero phonon line is observed at 861~nm with cavity modes decorating the phonon sideband at approximately 895~nm and 948~nm. Inset depicts Lorentzian fit of cavity mode with Q=2600. (b). 77 Kelvin photoluminescence spectrum of narrow V2 zero phonon line formed from a 0.6~$\mu$J UV pulse, with no observable h-site emission, measured from an (c.) irradiation site with highly-localized emission, where the k-site ZPL is present only within the diffraction limited spot. (d.) Characteristic k-site spin resonance observed in an adjacent cavity irradiated with a single 0.95~$\mu$J pulse.}
\label{Spect}
\end{figure*}

Fig.~\ref{Spect} presents results spanning the single-pulse pre-amorphization range. We measure PL after single-pulse laser irradiation at 77~K to obtain higher spectral resolution and confirm the presence of spin-defect ZPLs. Under high fluence, we observe ensemble emission possessing both the h- and k-site ZPLs, and at decreased irradiation fluence observe emission from either an h- or k-site ZPL. Bright h-site V$_{Si}^-$ emission is observed integrated in a cavity with quality factor of Q=2600 (Fig.~\ref{Spect}a). K-site emission is observed with a narrow V2 ZPL (Fig.~\ref{Spect}b) from a highly localized irradiation site (Fig.~\ref{Spect}c). The k-site ZPL is present only in a diffraction limited spot surrounding the irradiation site and spectroscopy reveals no h-site ZPLs, suggesting an isolated k-site emitter. The V1$^\prime$ dipole orientation is orthogonal to the c-axis and thus more readily excited than the off-axis dipole of the k-site V2; therefore if an h-site was present the emission would be evident in conditions where k-site is also observed.  We additionally note the calculated formation energies of h- and k-site monovacancy are approximately equal \cite{KuateDefo2018} and thus render the irradiation site unlikely to contain numerous k-site emitters with no h-site present. The narrow V2 ZPL linewidth further suggests a single emitter. In an attempt to determine the number of emitters, single photon correlation measurements are performed on the irradiation site of Fig.~\ref{Spect}c using a Hanbury-Brown-Twiss interferometer. These measurements, and their limitations, are further discussed in the Supplementary Information \cite{Day}. The spin signature of the k-site emitters formed across the energy window is investigated at room temperature via off-resonant optically detected magnetic resonance (ODMR). Fig.~\ref{Spect}d shows clear spin resonance at the k-site's zero-field splitting of 70 MHz. These findings provide evidence that this formation process preserves the spin and optical purity of the emitters, and the photonic mode of the cavities. 

\begin{figure*}
\includegraphics{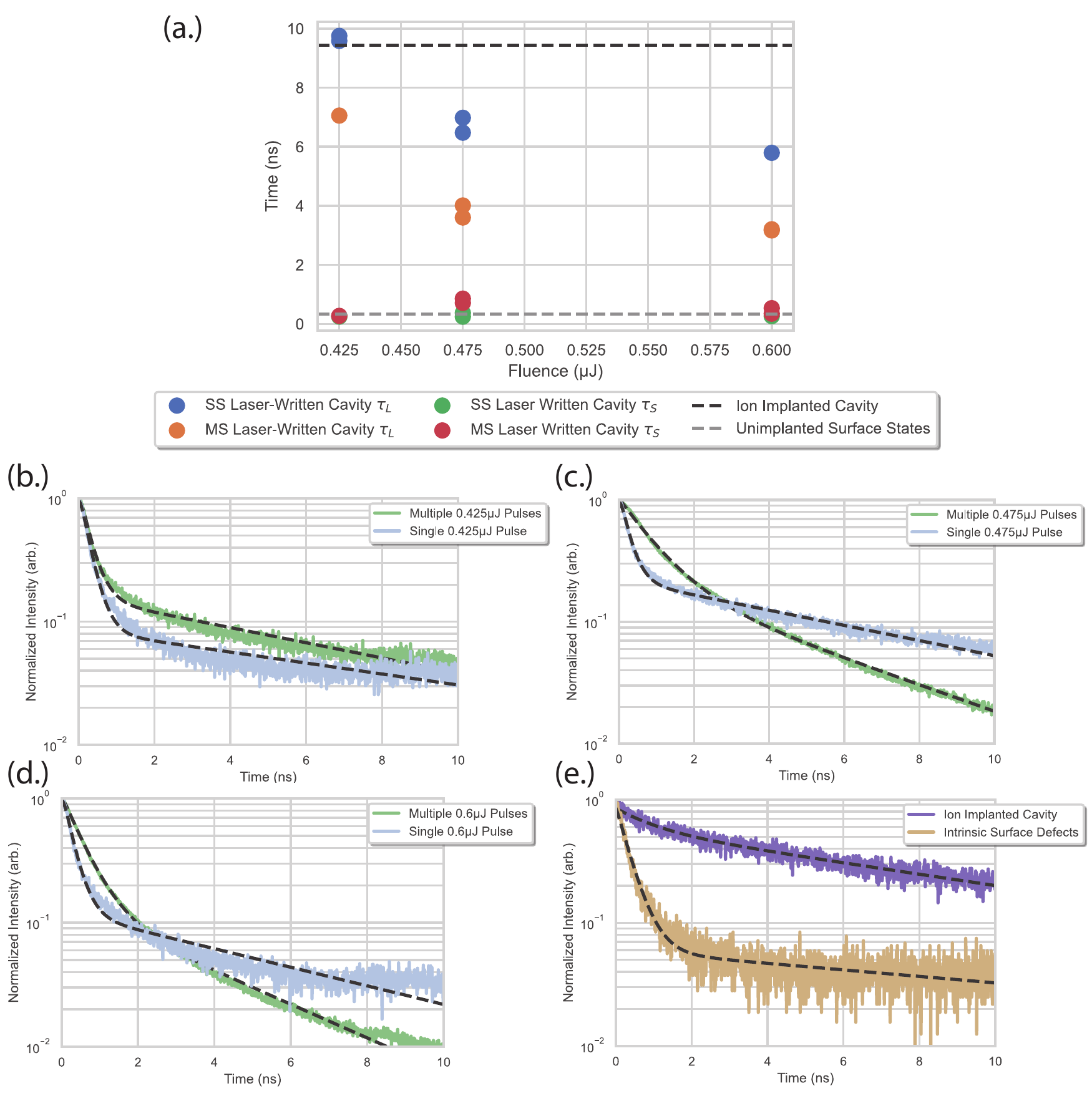}
\caption{\label{fig:wide} \textbf{Lifetime analysis of laser-irradiated cavity-integrated silicon monovacancy defects} (a.) Lifetimes plotted as a function of fluence, where SS and MS denote single-shot (1 UV pulse) and multi-shot (2-3 UV pulses), respectively. Horizontal bars depict the decay of an ion-implanted cavity-integrated sample used as a comparison metric (black), and an unimplanted cavity adjacent to the ones studied throughout the work (gray). Decay times are extracted from the fits of (b-e).}
\label{Lifetime}
\end{figure*}

The physical mechanisms employed in ion implantation, electron irradiation, previously-reported femto-second laser writing, and our above-bandgap nano-second laser writing are notably different, although an end goal is similar: local modification of the crystal structures to create high quality optically active spin defects. It would therefore be useful to evaluate the "collateral" creation of non-useful defects for a given technique. These considerations motivated our study of the {\it lifetimes} of the generated V$_{Si}^-$  and their correlation with irradiation fluence and pulse count. A summary of the results is shown in Fig.~\ref{Lifetime}.

A dose array is generated as a function of fluence across numerous unimplanted PCCs \cite{Day}, using a single irradiation pulse in one set of cavities, and multiple pulses in the other. The excited state lifetimes of emitters formed from single and multiple-pulse irradiation are plotted in Fig.~\ref{Lifetime}a;  the decay components are extracted from a fit of the transient data summarized in Fig.~\ref{Lifetime}b-e. Data taken from a neighboring unirradiated cavity, and from a reference cavity implanted with a density of 1$\times10^{12}$ carbon ions are also plotted, for comparison.

The transient decay data of Fig.~\ref{Lifetime}b-d reveal a significant short decay process that is comparable for all laser-irradiated samples, with a mean lifetime of 409~ps, insensitive to a change in irradiation fluence or pulse count. The short decay component might be related to {\it intrinsic defects} on the surface or inherent in the unprocessed cavities. As a comparison, we measure the lifetime in unimplanted, unirradiated cavities as shown in Fig.~\ref{Lifetime}e (light-brown). Indeed, the decay time of 345~ps found in the unimplanted cavity matches the short decay processes observed in the laser-irradiated cavities. We thus attribute this short-lived contribution to the surface defects. This observed contribution is consistent with the findings of Fig~\ref{Spect}, where the spin contrast of 0.2\% is diminished by background luminescence, relative to the 0.3-0.5\% typically observed in thermally unannealed ion-implanted near-surface emitters \cite{Dietz2022a}. This cavity nanoscope illuminates the presence of a short decay component attributable to intrinsic surface defects, either from the base material or the cavity fabrication. 

In contrast, the long decay components of Fig.~\ref{Lifetime}b-d corresponding to the excited state lifetime of V$_{Si}^-$ are strongly dependent upon the irradiation conditions, revealing an exponential reduction in the lifetime at increased irradiation fluence, and a uniform reduction in lifetime upon application of multiple pulses. Higher fluence and high pulse count may therefore increase non-radiative decay channels or create other unintended defects that quench the monovacancies' fluorescence. Therefore, we conclude from the trends evident in  Fig.~\ref{Lifetime}a that beneath the amorphization threshold of the cavities, lattice damage is induced as irradiation fluence and pulse count is increased--resulting in reduced emitter lifetime and therefore linewidth broadening.

\begin{figure}[ht!]
\includegraphics[scale=0.94]{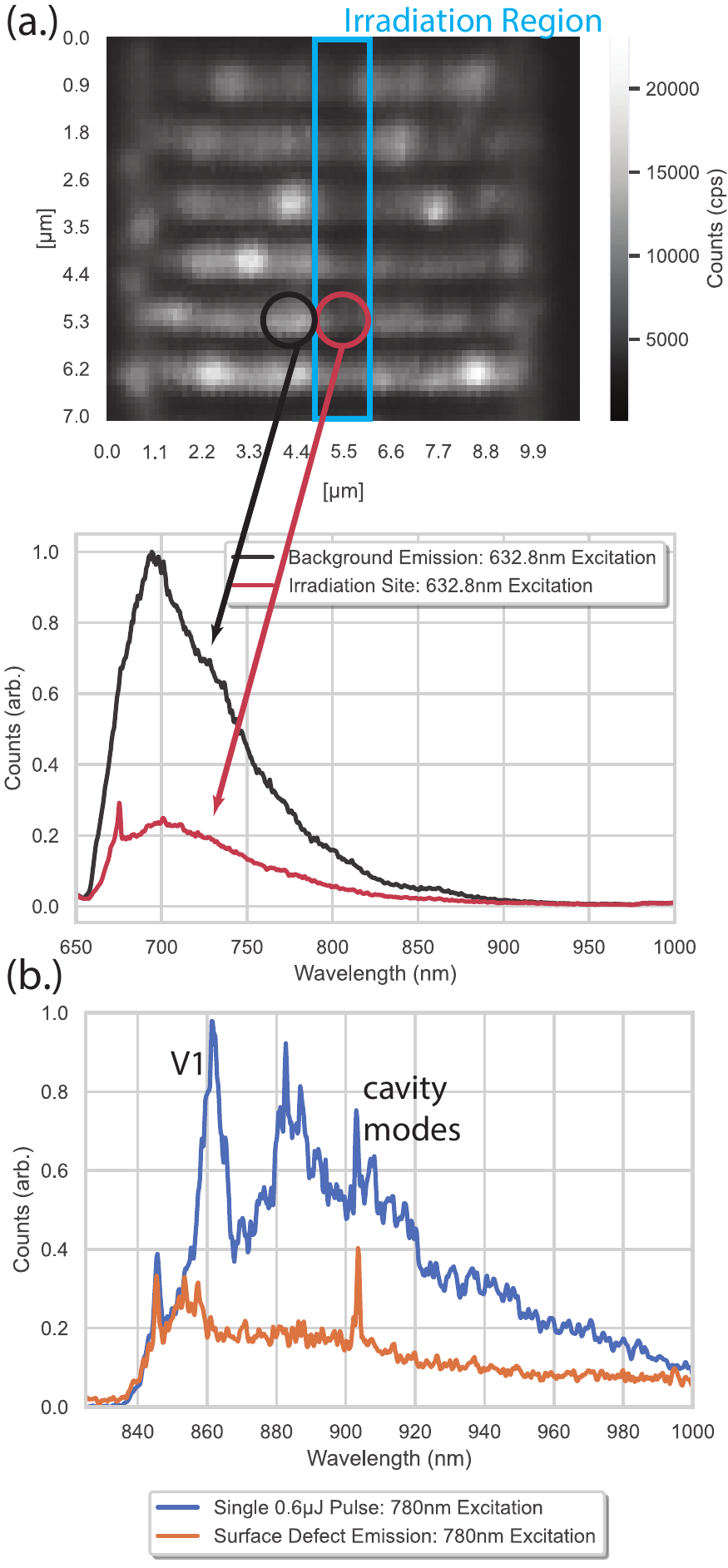}
\caption{\textbf{Single-shot UV laser annealing} (a.) Intensity distribution of surface state emission (white) from six cavities each irradiated by a single UV pulse (measured under CW 632.8~nm excitation). Blue box denotes UV irradiation region, revealing both an 80\% reduction in surface defect emission and emergence of a $C_{Si}V_C$ ZPL, where PL spectra is presented for the regions unirradiated (black) and irradiated (red) by a single UV pulse. (b.) Real-time in-situ monitoring of V$_{Si}^-$ formation before (orange) and after (blue) a single UV pulse, where surface defects weakly couple to the cavity modes prior to incorporating V$_{Si}^-$, revealing the negligible detriment of a subsequent single pulse to the cavities.}
\label{Surface}
\end{figure}

We have already shown that surface defects have an impact on the measured spin and optical characteristics of laser written cavities--these defects could also play a mediating role in V$_{Si}^-$ formation. Characterizing these states' evolution under laser irradiation is important in ultimately developing an accurate physical model of the laser writing process. We irradiate a single UV pulse on six cavities possessing pre-existing surface states (Fig.~\ref{Surface}a blue boxed region) and then measure PL in both irradiated and unirradiated regions using a 632.8~nm CW helium-neon laser. An avalanche photodiode raster scan reveals surface defect emission, which appears as bright regions along the unirradiated edges of the cavities (Fig.~4a). A typical spectrum is provided (black).

Surprisingly, the intrinsic defect PL is dimmer in the irradiated region. Comparing the PL spectra of the irradiated (red) region of the cavity where the k-site emission of Fig.~\ref{Spect}b was observed to the unimplanted surface defect emission (black), we note both an 80\% reduction in surface defect emission, and an emergent peak centered around 665~nm--commonly observed in carbon anti-site vacancy pair ($C_{Si}V_C$) emission associated with the 640-670~nm A-line ZPLs \cite{Dietz2022a, Castelletto2014}. $C_{Si}V_C$ has a lower calculated formation energy ($\sim$6~eV) than V$_{Si}^-$ ($\sim$7.5~eV) \cite{KuateDefo2018}. Thus, it is not surprising that laser irradiation conditions which produce V$_{Si}^-$ also give rise to the observed $C_{Si}V_C$; one mechanism may be the formation of $C_{Si}V_C$ pairs from pre-existing $C_{Si}$ and $V_C$.

Having characterized the surface defect emission we can leverage our understanding to aid in the {\it in-situ} study of V$_{Si}^-$ formation. Fig.~\ref{Surface}b depicts the cavity mode volume immediately before (orange) and after (blue) irradiation of a single 0.6$\mu$J UV pulse, while pumped by a 780~nm excitation laser. Prior to irradiation the surface state emission weakly couples to the cavity, enabling observation of the resonant modes. After irradiation with a single UV pulse, a clear V1 ZPL is observed with the same cavity modes still present--indicating the process which forms $V_{Si}^-$ within the cavity mode volume preserves the photonic properties of the device.

In the nano-second order pulsed laser regime, the dominant means of energy transfer to the lattice is thermal, mediated by Joule heating from the laser pulse \cite{Gattass2008}. Furthermore, above-bandgap excitation induces efficient energy transfer by the increased absorption constant. While the cavities are held at 77 Kelvin in the cryostat, thermal energy is thus rapidly deposited and dissipated in the diffraction-limited spot. We observe a striking similarity in the reduction of surface defect emission with commensurate observation of V$_{Si}^-$ and $C_{Si}V_C$ ZPLs in our process and previously-reported studies in which samples were annealed at 850$^\circ$C and quenched in water \cite{Dietz2022a}. We additionally note the potential of an above-bandgap recombination enhanced defect reaction annealing out the intrinsic surface defects, where 3.67eV photons promote electrons across 4H-SiC's bandgap, and generated electrons and holes can be captured by defect sites to mediate highly localized multi-phonon emission and subsequent defect diffusion \cite{Skowronski2002, Kimerling1978, Lang1974, Gadalla2021}. These results suggest a UV laser pulse could be driving a thermally-mediated defect reaction, raising the possibility that pre-existing defects may play a role in the laser-writing of V$_{Si}^-$ as well as $C_{Si}V_C$. This enables the possibility of utilizing the presented laser writing method to not only form the V$_{Si}^-$ defects of interest within the cavities, but also laser anneal unwanted intrinsic surface defects. Further work is nonetheless needed to explore optimal conditions for surface treatment and background defect annealing.

We have shown deterministic cavity integration of $V_{Si}^-$ using a new method of above-bandgap nano-second pulsed direct laser writing. The process preserves the spin and optical properties of the defect, and the cavity mode of the nanophotonic device, demonstrating narrow linewidths and lifetimes of near-surface spins longer than observed for defects formed in previous laser-writing demonstrations, where the defects were created microns below the surface in bulk material. Additionally, we have developed an understanding of the process which drives the defect formation by studying how the local environment of intrinsic surface defects shows local annealing. We have signatures supporting that our process forms single $V_{Si}^-$ by the count rate, spatial confinement, linewidth, and isolation of h-site and k-site ZPLs in varied irradiation sites. Unambiguous observation of single photon correlations is inhibited by the presence of surface defects, which could potentially be mitigated in future work by laser writing cavities on purer material, or utilizing surface-treatment methods to reduce background defect fluorescence. This method can be readily extended to other semiconductors such as Gallium Nitride, Silicon and 3C-SiC to further explore the defect formation mechanism. 

Localized incorporation of emitters in photonic structures with a high yield of single emitters is an outstanding challenge in advancing defects in solids for quantum networking. The incorporation of single emitters into prefabricated optical cavities presents a high throughput means of generating emitters and greatly eases the challenge of generating defects {\it ex-situ} (i.e. with an electron beam, ion beam or focused ion beam) followed by large scale characterization. The demonstrated technique allows not only selective formation of defects, but also rapid {\it in-situ} characterization of the defects that are formed in cavities, requiring only the inclusion of one more laser in the confocal system with no need for post-irradiation annealing to improve the emitter characteristics. This process realizes defect formation with single site addressing of nanophotonic cavity modes, offering a new tool in engineering optimal cavity-emitter coupling for system development in quantum networking and multi-emitter cavity-QED \cite{Lukin2022,Patton}.

\section{Methods}
\subsection{Sample Preparation}
The sample used in this study is an epitaxially-grown 100-200-100~nm p-i-p-n stack of 4H-SiC (Norstel AB), comprised of $10^{18}$ cm$^{-3}$ Aluminum-doped p-type, intrinsic, and $10^{19}$ cm$^{-3}$ Nitrogen-doped n-type bulk. Photonic crystal cavities are fabricated on the sample using the electron-beam lithography and reactive ion etching process given in \cite{Bracher2017,Dietz2022a}.

The ion-implanted cavity sample used as a reference in Fig.~\ref{Lifetime} had the same epitaxial conditions and cavity fabrication as described above, but was implanted with 1$\times10^{12}$ C$^{12}$ ions with an energy of 70keV at $7^{\circ}$ off c-axis to form an ensemble of V$_{Si}^-$ at a depth of 132~nm given by SRIM stopping range calculations \cite{Dietz2022a}. Photoluminescence spectra of characteristic V$_{Si}^-$ emission measured from this sample are provided in the supplement \cite{Day}. 
\subsection{Defect Formation via Pulsed Laser Writing}
A SpectraPhysics nanosecond-pulsed nitrogen laser is used to direct individual pulses of 4~ns 337.1~nm laser light onto unimplanted fabricated photonic crystal cavities. The pulse fluence is controlled with a variable neutral density attenuator. Pulses are focused onto the sample with an Olympus 40x 0.6 NA correction-collar objective. 

\subsection{Spectroscopy}
Spectroscopy is performed with an Acton 500i spectrograph using a 150 gr/mm 800~nm Blaze grating and a Princeton Instruments visible 1340x400 BI camera. For Fig.~\ref{SystemOverview}c, Fig.~ \ref{Spect}a-b and Fig.~\ref{Surface}b, following irradiation from the above-bandgap laser-writing pulse, a Coherent Mira900 Titanium Sapphire laser tuned to a wavelength of 780~nm is used to excite off-resonant photoluminescence of the defects formed within the cavity volume. The excitation laser intensity is controlled with a rotatable half-wave plate and polarizing beam splitter. The laser is filtered with a 780~nm band-pass filter. Defect emission is filtered with an 850~nm long-pass filter and collected with a 10x 0.28 NA Mitutoyo objective into a 25$\mu$m core multi-mode fiber. Fig.~ \ref{Spect}c uses a single-mode fiber for improved scan resolution. 

For Fig.~\ref{Surface}, 1~mW of a Melles-Griot 05-LHP-991 632.8~nm helium-neon laser is used to efficiently excite surface defects and $C_{Si}V_C$. The excitation is filtered with a 632.8~nm band pass filter, and the collection is filtered with a 650~nm long pass filter. 

\subsection{Excited State Lifetime}
Excited state lifetime is measured with the same optical system as described in Methods Section C, but with an excitation power of 80$~\mu$W, a zero-phonon-line 900~nm short-pass filter added to the emission collection, and an Excelitas SPCM-AQRH-14-FC avalanche photodiode (APD) to detect photon counts. The lifetime florescence software PicoHarp is used to measure the decay from the APD counts. The results of Fig.~\ref{Lifetime} are also summarized in Table~\ref{tab:table1}. 

\begin{table}[h]
\caption{\label{tab:table1}%
Summary of excited state lifetime decay components of Fig.~\ref{Lifetime}. SS and MS denote cavities irradiated with a single shot or multiple shots of UV irradiation, respectively.}
\begin{ruledtabular}
\begin{tabular}{ccc}
\textrm{Fluence [$\mu$J/Pulse]}&
\textrm{SS UV Decay [ns]}&
\textrm{MS UV Decay [ns]}\\
\colrule
0.425 & 9.66$\pm$0.08 & 7.05\\
0.475 & 6.73$\pm$0.25 & 3.81$\pm$0.20 \\
0.600 & 5.79 & 3.18$\pm$0.02 \\
\end{tabular}
\begin{tabular}{cc}
\textrm{Measurement}&
\textrm{Decay [ns]}\\
\colrule
Ion Implanted Cavity & 9.72$\pm$0.28 \\
Surface States & 0.345$\pm$0.01 \\
Laser Writing Short Decay Component & 0.409 (avg.)
\end{tabular}
\end{ruledtabular}
\end{table} 

The raw decay data is fit against the convolution of a double exponential and a gaussian window (to model the APD instrument response function (IRF), where an IRF width of $\sim$300ps is consistently extracted from the decays), with short and long decay times found from a fit. The bi-exponential is necessary to capture the transient effects of the surface defects discussed throughout the work. The equation used to generate the fit is given by: 

\begin{equation}
    y(t) = (e^{-(t-\mu)^2 / 2\sigma^2})\ast(Ae^{-t/\tau_s} + Be^{-t/\tau_L})
\end{equation}

\subsection{Optically Detected Magnetic Resonance}
A home-built confocal microscope is used to perform off-resonant ODMR using an external-cavity diode-laser tuned to 865~nm with 1~mW of input power. Detailed system schematic can be found in \cite{Dietz2022}.

\section*{Author contributions}
A.M.D., J.R.D., and E.L.H. designed the research. A.M.D. and J.R.D. performed all measurements and data analysis, with assistance from M.S. and M.Y. All authors contributed to preparing the manuscript. 

\section*{Acknowledgments}
We thank Dr. Xingyu Zhang for fabrication support. This work was supported by the Science and Technology Center for Integrated Quantum Materials, NSF Grant No. DMR-1231319. Portions of this work were performed at the Harvard University Center for Nanoscale Systems (CNS); a member of the National Nanotechnology Coordinated Infrastructure Network (NNCI), which is supported by the National Science Foundation under NSF award no. ECCS-2025158. J.R.D acknowledges funding from NSF RAISE-TAQS Award 1839164. M.S. acknowledges funding from a NASA Space Technology Graduate Research Fellowship. M.Y. acknowledges funding from the Department of Defense (DoD) through the National Defense Science and Engineering Graduate (NDSEG) Fellowship Program.

\section*{Data Availability}
The data that support the findings of the work are available from the corresponding author upon reasonable request.

\bibliography{LWref.bib}

\begin{thebibliography}{30}%
\makeatletter
\providecommand \@ifxundefined [1]{%
 \@ifx{#1\undefined}
}%
\providecommand \@ifnum [1]{%
 \ifnum #1\expandafter \@firstoftwo
 \else \expandafter \@secondoftwo
 \fi
}%
\providecommand \@ifx [1]{%
 \ifx #1\expandafter \@firstoftwo
 \else \expandafter \@secondoftwo
 \fi
}%
\providecommand \natexlab [1]{#1}%
\providecommand \enquote  [1]{``#1''}%
\providecommand \bibnamefont  [1]{#1}%
\providecommand \bibfnamefont [1]{#1}%
\providecommand \citenamefont [1]{#1}%
\providecommand \href@noop [0]{\@secondoftwo}%
\providecommand \href [0]{\begingroup \@sanitize@url \@href}%
\providecommand \@href[1]{\@@startlink{#1}\@@href}%
\providecommand \@@href[1]{\endgroup#1\@@endlink}%
\providecommand \@sanitize@url [0]{\catcode `\\12\catcode `\$12\catcode
  `\&12\catcode `\#12\catcode `\^12\catcode `\_12\catcode `\%12\relax}%
\providecommand \@@startlink[1]{}%
\providecommand \@@endlink[0]{}%
\providecommand \url  [0]{\begingroup\@sanitize@url \@url }%
\providecommand \@url [1]{\endgroup\@href {#1}{\urlprefix }}%
\providecommand \urlprefix  [0]{URL }%
\providecommand \Eprint [0]{\href }%
\providecommand \doibase [0]{https://doi.org/}%
\providecommand \selectlanguage [0]{\@gobble}%
\providecommand \bibinfo  [0]{\@secondoftwo}%
\providecommand \bibfield  [0]{\@secondoftwo}%
\providecommand \translation [1]{[#1]}%
\providecommand \BibitemOpen [0]{}%
\providecommand \bibitemStop [0]{}%
\providecommand \bibitemNoStop [0]{.\EOS\space}%
\providecommand \EOS [0]{\spacefactor3000\relax}%
\providecommand \BibitemShut  [1]{\csname bibitem#1\endcsname}%
\let\auto@bib@innerbib\@empty
\bibitem [{\citenamefont {Stas}\ \emph {et~al.}(2022)\citenamefont {Stas},
  \citenamefont {Huan}, \citenamefont {Machielse}, \citenamefont {Knall},
  \citenamefont {Suleymanzade}, \citenamefont {Pingault}, \citenamefont
  {Sutula}, \citenamefont {Ding}, \citenamefont {Knaut}, \citenamefont
  {Assumpcao}, \citenamefont {Wei}, \citenamefont {Bhaskar}, \citenamefont
  {Riedinger}, \citenamefont {Sukachev}, \citenamefont {Park}, \citenamefont
  {Lon{\v{c}}ar}, \citenamefont {Levonian},\ and\ \citenamefont
  {Lukin}}]{Stas2022}%
  \BibitemOpen
  \bibfield  {author} {\bibinfo {author} {\bibfnamefont {P.-J.}\ \bibnamefont
  {Stas}}, \bibinfo {author} {\bibfnamefont {Y.~Q.}\ \bibnamefont {Huan}},
  \bibinfo {author} {\bibfnamefont {B.}~\bibnamefont {Machielse}}, \bibinfo
  {author} {\bibfnamefont {E.~N.}\ \bibnamefont {Knall}}, \bibinfo {author}
  {\bibfnamefont {A.}~\bibnamefont {Suleymanzade}}, \bibinfo {author}
  {\bibfnamefont {B.}~\bibnamefont {Pingault}}, \bibinfo {author}
  {\bibfnamefont {M.}~\bibnamefont {Sutula}}, \bibinfo {author} {\bibfnamefont
  {S.~W.}\ \bibnamefont {Ding}}, \bibinfo {author} {\bibfnamefont {C.~M.}\
  \bibnamefont {Knaut}}, \bibinfo {author} {\bibfnamefont {D.~R.}\ \bibnamefont
  {Assumpcao}}, \bibinfo {author} {\bibfnamefont {Y.-C.}\ \bibnamefont {Wei}},
  \bibinfo {author} {\bibfnamefont {M.~K.}\ \bibnamefont {Bhaskar}}, \bibinfo
  {author} {\bibfnamefont {R.}~\bibnamefont {Riedinger}}, \bibinfo {author}
  {\bibfnamefont {D.~D.}\ \bibnamefont {Sukachev}}, \bibinfo {author}
  {\bibfnamefont {H.}~\bibnamefont {Park}}, \bibinfo {author} {\bibfnamefont
  {M.}~\bibnamefont {Lon{\v{c}}ar}}, \bibinfo {author} {\bibfnamefont {D.~S.}\
  \bibnamefont {Levonian}},\ and\ \bibinfo {author} {\bibfnamefont {M.~D.}\
  \bibnamefont {Lukin}},\ }\bibfield  {title} {\bibinfo {title} {{Robust
  multi-qubit quantum network node with integrated error detection}},\
  }\bibfield  {journal} {\bibinfo  {journal} {arXiv:2207.13128}\ }\href
  {https://doi.org/10.48550/arxiv.2207.13128} {10.48550/arxiv.2207.13128}
  (\bibinfo {year} {2022}),\ \Eprint {https://arxiv.org/abs/2207.13128}
  {arXiv:2207.13128} \BibitemShut {NoStop}%
\bibitem [{\citenamefont {Bhaskar}\ \emph {et~al.}(2020)\citenamefont
  {Bhaskar}, \citenamefont {Riedinger}, \citenamefont {Machielse},
  \citenamefont {Levonian}, \citenamefont {Nguyen}, \citenamefont {Knall},
  \citenamefont {Park}, \citenamefont {Englund}, \citenamefont {Lon{\v{c}}ar},
  \citenamefont {Sukachev},\ and\ \citenamefont {Lukin}}]{Bhaskar2020}%
  \BibitemOpen
  \bibfield  {author} {\bibinfo {author} {\bibfnamefont {M.~K.}\ \bibnamefont
  {Bhaskar}}, \bibinfo {author} {\bibfnamefont {R.}~\bibnamefont {Riedinger}},
  \bibinfo {author} {\bibfnamefont {B.}~\bibnamefont {Machielse}}, \bibinfo
  {author} {\bibfnamefont {D.~S.}\ \bibnamefont {Levonian}}, \bibinfo {author}
  {\bibfnamefont {C.~T.}\ \bibnamefont {Nguyen}}, \bibinfo {author}
  {\bibfnamefont {E.~N.}\ \bibnamefont {Knall}}, \bibinfo {author}
  {\bibfnamefont {H.}~\bibnamefont {Park}}, \bibinfo {author} {\bibfnamefont
  {D.}~\bibnamefont {Englund}}, \bibinfo {author} {\bibfnamefont
  {M.}~\bibnamefont {Lon{\v{c}}ar}}, \bibinfo {author} {\bibfnamefont {D.~D.}\
  \bibnamefont {Sukachev}},\ and\ \bibinfo {author} {\bibfnamefont {M.~D.}\
  \bibnamefont {Lukin}},\ }\bibfield  {title} {\bibinfo {title} {{Experimental
  demonstration of memory-enhanced quantum communication}},\ }\href
  {https://doi.org/10.1038/s41586-020-2103-5} {\bibfield  {journal} {\bibinfo
  {journal} {Nature}\ }\textbf {\bibinfo {volume} {580}},\ \bibinfo {pages}
  {60} (\bibinfo {year} {2020})},\ \Eprint {https://arxiv.org/abs/1909.01323}
  {arXiv:1909.01323} \BibitemShut {NoStop}%
\bibitem [{\citenamefont {Babin}\ \emph {et~al.}(2022)\citenamefont {Babin},
  \citenamefont {St{\"{o}}hr}, \citenamefont {Morioka}, \citenamefont
  {Linkewitz}, \citenamefont {Steidl}, \citenamefont {W{\"{o}}rnle},
  \citenamefont {Liu}, \citenamefont {Hesselmeier}, \citenamefont {Vorobyov},
  \citenamefont {Denisenko}, \citenamefont {Hentschel}, \citenamefont {Gobert},
  \citenamefont {Berwian}, \citenamefont {Astakhov}, \citenamefont {Knolle},
  \citenamefont {Majety}, \citenamefont {Saha}, \citenamefont {Radulaski},
  \citenamefont {Son}, \citenamefont {Ul-Hassan}, \citenamefont {Kaiser},\ and\
  \citenamefont {Wrachtrup}}]{Babin2022}%
  \BibitemOpen
  \bibfield  {author} {\bibinfo {author} {\bibfnamefont {C.}~\bibnamefont
  {Babin}}, \bibinfo {author} {\bibfnamefont {R.}~\bibnamefont {St{\"{o}}hr}},
  \bibinfo {author} {\bibfnamefont {N.}~\bibnamefont {Morioka}}, \bibinfo
  {author} {\bibfnamefont {T.}~\bibnamefont {Linkewitz}}, \bibinfo {author}
  {\bibfnamefont {T.}~\bibnamefont {Steidl}}, \bibinfo {author} {\bibfnamefont
  {R.}~\bibnamefont {W{\"{o}}rnle}}, \bibinfo {author} {\bibfnamefont
  {D.}~\bibnamefont {Liu}}, \bibinfo {author} {\bibfnamefont {E.}~\bibnamefont
  {Hesselmeier}}, \bibinfo {author} {\bibfnamefont {V.}~\bibnamefont
  {Vorobyov}}, \bibinfo {author} {\bibfnamefont {A.}~\bibnamefont {Denisenko}},
  \bibinfo {author} {\bibfnamefont {M.}~\bibnamefont {Hentschel}}, \bibinfo
  {author} {\bibfnamefont {C.}~\bibnamefont {Gobert}}, \bibinfo {author}
  {\bibfnamefont {P.}~\bibnamefont {Berwian}}, \bibinfo {author} {\bibfnamefont
  {G.~V.}\ \bibnamefont {Astakhov}}, \bibinfo {author} {\bibfnamefont
  {W.}~\bibnamefont {Knolle}}, \bibinfo {author} {\bibfnamefont
  {S.}~\bibnamefont {Majety}}, \bibinfo {author} {\bibfnamefont
  {P.}~\bibnamefont {Saha}}, \bibinfo {author} {\bibfnamefont {M.}~\bibnamefont
  {Radulaski}}, \bibinfo {author} {\bibfnamefont {N.~T.}\ \bibnamefont {Son}},
  \bibinfo {author} {\bibfnamefont {J.}~\bibnamefont {Ul-Hassan}}, \bibinfo
  {author} {\bibfnamefont {F.}~\bibnamefont {Kaiser}},\ and\ \bibinfo {author}
  {\bibfnamefont {J.}~\bibnamefont {Wrachtrup}},\ }\bibfield  {title} {\bibinfo
  {title} {{Fabrication and nanophotonic waveguide integration of silicon
  carbide colour centres with preserved spin-optical coherence}},\ }\href
  {https://doi.org/10.1038/S41563-021-01148-3} {\bibfield  {journal} {\bibinfo
  {journal} {Nature Materials}\ }\textbf {\bibinfo {volume} {21}},\ \bibinfo
  {pages} {67} (\bibinfo {year} {2022})}\BibitemShut {NoStop}%
\bibitem [{\citenamefont {Lukin}\ \emph
  {et~al.}(2020{\natexlab{a}})\citenamefont {Lukin}, \citenamefont {Dory},
  \citenamefont {Guidry}, \citenamefont {Yang}, \citenamefont {Mishra},
  \citenamefont {Trivedi}, \citenamefont {Radulaski}, \citenamefont {Sun},
  \citenamefont {Vercruysse}, \citenamefont {Ahn},\ and\ \citenamefont
  {Vu{\v{c}}kovi{\'{c}}}}]{Lukin2020a}%
  \BibitemOpen
  \bibfield  {author} {\bibinfo {author} {\bibfnamefont {D.~M.}\ \bibnamefont
  {Lukin}}, \bibinfo {author} {\bibfnamefont {C.}~\bibnamefont {Dory}},
  \bibinfo {author} {\bibfnamefont {M.~A.}\ \bibnamefont {Guidry}}, \bibinfo
  {author} {\bibfnamefont {K.~Y.}\ \bibnamefont {Yang}}, \bibinfo {author}
  {\bibfnamefont {S.~D.}\ \bibnamefont {Mishra}}, \bibinfo {author}
  {\bibfnamefont {R.}~\bibnamefont {Trivedi}}, \bibinfo {author} {\bibfnamefont
  {M.}~\bibnamefont {Radulaski}}, \bibinfo {author} {\bibfnamefont
  {S.}~\bibnamefont {Sun}}, \bibinfo {author} {\bibfnamefont {D.}~\bibnamefont
  {Vercruysse}}, \bibinfo {author} {\bibfnamefont {G.~H.}\ \bibnamefont
  {Ahn}},\ and\ \bibinfo {author} {\bibfnamefont {J.}~\bibnamefont
  {Vu{\v{c}}kovi{\'{c}}}},\ }\bibfield  {title} {\bibinfo {title}
  {{4H-silicon-carbide-on-insulator for integrated quantum and nonlinear
  photonics}},\ }\href {https://doi.org/10.1038/S41566-019-0556-6} {\bibfield
  {journal} {\bibinfo  {journal} {Nature Photonics}\ }\textbf {\bibinfo
  {volume} {14}},\ \bibinfo {pages} {330} (\bibinfo {year}
  {2020}{\natexlab{a}})}\BibitemShut {NoStop}%
\bibitem [{\citenamefont {Bracher}\ \emph {et~al.}(2017)\citenamefont
  {Bracher}, \citenamefont {Zhang},\ and\ \citenamefont {Hu}}]{Bracher2017}%
  \BibitemOpen
  \bibfield  {author} {\bibinfo {author} {\bibfnamefont {D.~O.}\ \bibnamefont
  {Bracher}}, \bibinfo {author} {\bibfnamefont {X.}~\bibnamefont {Zhang}},\
  and\ \bibinfo {author} {\bibfnamefont {E.~L.}\ \bibnamefont {Hu}},\
  }\bibfield  {title} {\bibinfo {title} {{Selective Purcell enhancement of two
  closely linked zero-phonon transitions of a silicon carbide color center}},\
  }\href {https://doi.org/10.1073/PNAS.1704219114} {\bibfield  {journal}
  {\bibinfo  {journal} {Proceedings of the National Academy of Sciences}\
  }\textbf {\bibinfo {volume} {114}},\ \bibinfo {pages} {4060} (\bibinfo {year}
  {2017})},\ \Eprint {https://arxiv.org/abs/1609.03918} {arXiv:1609.03918}
  \BibitemShut {NoStop}%
\bibitem [{\citenamefont {Crook}\ \emph {et~al.}(2020)\citenamefont {Crook},
  \citenamefont {Anderson}, \citenamefont {Miao}, \citenamefont {Bourassa},
  \citenamefont {Lee}, \citenamefont {Bayliss}, \citenamefont {Bracher},
  \citenamefont {Zhang}, \citenamefont {Abe}, \citenamefont {Ohshima},
  \citenamefont {Hu},\ and\ \citenamefont {Awschalom}}]{Crook2020}%
  \BibitemOpen
  \bibfield  {author} {\bibinfo {author} {\bibfnamefont {A.~L.}\ \bibnamefont
  {Crook}}, \bibinfo {author} {\bibfnamefont {C.~P.}\ \bibnamefont {Anderson}},
  \bibinfo {author} {\bibfnamefont {K.~C.}\ \bibnamefont {Miao}}, \bibinfo
  {author} {\bibfnamefont {A.}~\bibnamefont {Bourassa}}, \bibinfo {author}
  {\bibfnamefont {H.}~\bibnamefont {Lee}}, \bibinfo {author} {\bibfnamefont
  {S.~L.}\ \bibnamefont {Bayliss}}, \bibinfo {author} {\bibfnamefont {D.~O.}\
  \bibnamefont {Bracher}}, \bibinfo {author} {\bibfnamefont {X.}~\bibnamefont
  {Zhang}}, \bibinfo {author} {\bibfnamefont {H.}~\bibnamefont {Abe}}, \bibinfo
  {author} {\bibfnamefont {T.}~\bibnamefont {Ohshima}}, \bibinfo {author}
  {\bibfnamefont {E.~L.}\ \bibnamefont {Hu}},\ and\ \bibinfo {author}
  {\bibfnamefont {D.~D.}\ \bibnamefont {Awschalom}},\ }\bibfield  {title}
  {\bibinfo {title} {{Purcell enhancement of a single silicon carbide color
  center with coherent spin control}},\ }\href
  {https://doi.org/10.1021/ACS.NANOLETT.0C00339} {\bibfield  {journal}
  {\bibinfo  {journal} {Nano Letters}\ }\textbf {\bibinfo {volume} {20}},\
  \bibinfo {pages} {3427} (\bibinfo {year} {2020})},\ \Eprint
  {https://arxiv.org/abs/2003.00042} {arXiv:2003.00042} \BibitemShut {NoStop}%
\bibitem [{\citenamefont {Gadalla}\ \emph {et~al.}(2021)\citenamefont
  {Gadalla}, \citenamefont {Greenspon}, \citenamefont {Defo}, \citenamefont
  {Zhang},\ and\ \citenamefont {Hu}}]{Gadalla2021}%
  \BibitemOpen
  \bibfield  {author} {\bibinfo {author} {\bibfnamefont {M.~N.}\ \bibnamefont
  {Gadalla}}, \bibinfo {author} {\bibfnamefont {A.~S.}\ \bibnamefont
  {Greenspon}}, \bibinfo {author} {\bibfnamefont {R.~K.}\ \bibnamefont {Defo}},
  \bibinfo {author} {\bibfnamefont {X.}~\bibnamefont {Zhang}},\ and\ \bibinfo
  {author} {\bibfnamefont {E.~L.}\ \bibnamefont {Hu}},\ }\bibfield  {title}
  {\bibinfo {title} {{Enhanced cavity coupling to silicon vacancies in 4H
  silicon carbide using laser irradiation and thermal annealing}},\ }\bibfield
  {journal} {\bibinfo  {journal} {Proceedings of the National Academy of
  Sciences of the United States of America}\ }\textbf {\bibinfo {volume}
  {118}},\ \href {https://doi.org/10.1073/PNAS.2021768118}
  {10.1073/PNAS.2021768118} (\bibinfo {year} {2021})\BibitemShut {NoStop}%
\bibitem [{\citenamefont {Knall}\ \emph {et~al.}(2022)\citenamefont {Knall},
  \citenamefont {Knaut}, \citenamefont {Bekenstein}, \citenamefont {Assumpcao},
  \citenamefont {Stroganov}, \citenamefont {Gong}, \citenamefont {Huan},
  \citenamefont {Stas}, \citenamefont {Machielse}, \citenamefont {Chalupnik},
  \citenamefont {Levonian}, \citenamefont {Suleymanzade}, \citenamefont
  {Riedinger}, \citenamefont {Park}, \citenamefont {Lon{\v{c}}ar},
  \citenamefont {Bhaskar},\ and\ \citenamefont {Lukin}}]{Knall2022}%
  \BibitemOpen
  \bibfield  {author} {\bibinfo {author} {\bibfnamefont {E.~N.}\ \bibnamefont
  {Knall}}, \bibinfo {author} {\bibfnamefont {C.~M.}\ \bibnamefont {Knaut}},
  \bibinfo {author} {\bibfnamefont {R.}~\bibnamefont {Bekenstein}}, \bibinfo
  {author} {\bibfnamefont {D.~R.}\ \bibnamefont {Assumpcao}}, \bibinfo {author}
  {\bibfnamefont {P.~L.}\ \bibnamefont {Stroganov}}, \bibinfo {author}
  {\bibfnamefont {W.}~\bibnamefont {Gong}}, \bibinfo {author} {\bibfnamefont
  {Y.~Q.}\ \bibnamefont {Huan}}, \bibinfo {author} {\bibfnamefont {P.-J.}\
  \bibnamefont {Stas}}, \bibinfo {author} {\bibfnamefont {B.}~\bibnamefont
  {Machielse}}, \bibinfo {author} {\bibfnamefont {M.}~\bibnamefont
  {Chalupnik}}, \bibinfo {author} {\bibfnamefont {D.}~\bibnamefont {Levonian}},
  \bibinfo {author} {\bibfnamefont {A.}~\bibnamefont {Suleymanzade}}, \bibinfo
  {author} {\bibfnamefont {R.}~\bibnamefont {Riedinger}}, \bibinfo {author}
  {\bibfnamefont {H.}~\bibnamefont {Park}}, \bibinfo {author} {\bibfnamefont
  {M.}~\bibnamefont {Lon{\v{c}}ar}}, \bibinfo {author} {\bibfnamefont {M.~K.}\
  \bibnamefont {Bhaskar}},\ and\ \bibinfo {author} {\bibfnamefont {M.~D.}\
  \bibnamefont {Lukin}},\ }\bibfield  {title} {\bibinfo {title} {{Efficient
  Source of Shaped Single Photons Based on an Integrated Diamond Nanophotonic
  System}},\ }\href {https://doi.org/10.1103/PHYSREVLETT.129.053603} {\bibfield
   {journal} {\bibinfo  {journal} {Physical Review Letters}\ }\textbf {\bibinfo
  {volume} {129}},\ \bibinfo {pages} {053603} (\bibinfo {year}
  {2022})}\BibitemShut {NoStop}%
\bibitem [{\citenamefont {Sipahigil}\ \emph {et~al.}(2016)\citenamefont
  {Sipahigil}, \citenamefont {Evans}, \citenamefont {Sukachev}, \citenamefont
  {Burek}, \citenamefont {Borregaard}, \citenamefont {Bhaskar}, \citenamefont
  {Nguyen}, \citenamefont {Pacheco}, \citenamefont {Atikian}, \citenamefont
  {Meuwly}, \citenamefont {Camacho}, \citenamefont {Jelezko}, \citenamefont
  {Bielejec}, \citenamefont {Park}, \citenamefont {Lon{\v{c}}ar},\ and\
  \citenamefont {Lukin}}]{Sipahigil2016}%
  \BibitemOpen
  \bibfield  {author} {\bibinfo {author} {\bibfnamefont {A.}~\bibnamefont
  {Sipahigil}}, \bibinfo {author} {\bibfnamefont {R.~E.}\ \bibnamefont
  {Evans}}, \bibinfo {author} {\bibfnamefont {D.~D.}\ \bibnamefont {Sukachev}},
  \bibinfo {author} {\bibfnamefont {M.~J.}\ \bibnamefont {Burek}}, \bibinfo
  {author} {\bibfnamefont {J.}~\bibnamefont {Borregaard}}, \bibinfo {author}
  {\bibfnamefont {M.~K.}\ \bibnamefont {Bhaskar}}, \bibinfo {author}
  {\bibfnamefont {C.~T.}\ \bibnamefont {Nguyen}}, \bibinfo {author}
  {\bibfnamefont {J.~L.}\ \bibnamefont {Pacheco}}, \bibinfo {author}
  {\bibfnamefont {H.~A.}\ \bibnamefont {Atikian}}, \bibinfo {author}
  {\bibfnamefont {C.}~\bibnamefont {Meuwly}}, \bibinfo {author} {\bibfnamefont
  {R.~M.}\ \bibnamefont {Camacho}}, \bibinfo {author} {\bibfnamefont
  {F.}~\bibnamefont {Jelezko}}, \bibinfo {author} {\bibfnamefont
  {E.}~\bibnamefont {Bielejec}}, \bibinfo {author} {\bibfnamefont
  {H.}~\bibnamefont {Park}}, \bibinfo {author} {\bibfnamefont {M.}~\bibnamefont
  {Lon{\v{c}}ar}},\ and\ \bibinfo {author} {\bibfnamefont {M.~D.}\ \bibnamefont
  {Lukin}},\ }\bibfield  {title} {\bibinfo {title} {{An integrated diamond
  nanophotonics platform for quantum-optical networks}},\ }\href
  {https://doi.org/10.1126/SCIENCE.AAH6875/SUPPL_FILE/SIPAHIGIL-SM.PDF}
  {\bibfield  {journal} {\bibinfo  {journal} {Science}\ }\textbf {\bibinfo
  {volume} {354}},\ \bibinfo {pages} {847} (\bibinfo {year}
  {2016})}\BibitemShut {NoStop}%
\bibitem [{\citenamefont {Lukin}\ \emph {et~al.}(2022)\citenamefont {Lukin},
  \citenamefont {Guidry}, \citenamefont {Yang}, \citenamefont {Ghezellou},
  \citenamefont {Mishra}, \citenamefont {Abe}, \citenamefont {Ohshima},
  \citenamefont {Ul-Hassan},\ and\ \citenamefont {Vu{\v{c}}kovic}}]{Lukin2022}%
  \BibitemOpen
  \bibfield  {author} {\bibinfo {author} {\bibfnamefont {D.~M.}\ \bibnamefont
  {Lukin}}, \bibinfo {author} {\bibfnamefont {M.~A.}\ \bibnamefont {Guidry}},
  \bibinfo {author} {\bibfnamefont {J.}~\bibnamefont {Yang}}, \bibinfo {author}
  {\bibfnamefont {M.}~\bibnamefont {Ghezellou}}, \bibinfo {author}
  {\bibfnamefont {S.~D.}\ \bibnamefont {Mishra}}, \bibinfo {author}
  {\bibfnamefont {H.}~\bibnamefont {Abe}}, \bibinfo {author} {\bibfnamefont
  {T.}~\bibnamefont {Ohshima}}, \bibinfo {author} {\bibfnamefont
  {J.}~\bibnamefont {Ul-Hassan}},\ and\ \bibinfo {author} {\bibfnamefont
  {J.}~\bibnamefont {Vu{\v{c}}kovic}},\ }\bibfield  {title} {\bibinfo {title}
  {{Optical superradiance of a pair of color centers in an integrated
  silicon-carbide-on-insulator microresonator}},\ }\href@noop {} {\bibfield
  {journal} {\bibinfo  {journal} {arXiv 2202.04845}\ } (\bibinfo {year}
  {2022})},\ \Eprint {https://arxiv.org/abs/2202.04845v1} {arXiv:2202.04845v1}
  \BibitemShut {NoStop}%
\bibitem [{\citenamefont {Patton}\ \emph {et~al.}()\citenamefont {Patton},
  \citenamefont {Norman}, \citenamefont {Scalettar},\ and\ \citenamefont
  {Radulaski}}]{Patton}%
  \BibitemOpen
  \bibfield  {author} {\bibinfo {author} {\bibfnamefont {J.}~\bibnamefont
  {Patton}}, \bibinfo {author} {\bibfnamefont {V.~A.}\ \bibnamefont {Norman}},
  \bibinfo {author} {\bibfnamefont {R.~T.}\ \bibnamefont {Scalettar}},\ and\
  \bibinfo {author} {\bibfnamefont {M.}~\bibnamefont {Radulaski}},\ }\bibfield
  {title} {\bibinfo {title} {{All-Photonic Quantum Simulators with Spectrally
  Disordered Emitters}},\ }\href@noop {} {\bibfield  {journal} {\bibinfo
  {journal} {arXiv 2112.15469}\ }}\Eprint {https://arxiv.org/abs/2112.15469v1}
  {arXiv:2112.15469v1} \BibitemShut {NoStop}%
\bibitem [{\citenamefont {He}\ \emph {et~al.}()\citenamefont {He},
  \citenamefont {Li}, \citenamefont {Wen}, \citenamefont {Zhou}, \citenamefont
  {Lin}, \citenamefont {Hao}, \citenamefont {Xu}, \citenamefont {Li},\ and\
  \citenamefont {Guo}}]{FIB2022}%
  \BibitemOpen
  \bibfield  {author} {\bibinfo {author} {\bibfnamefont {Z.-X.}\ \bibnamefont
  {He}}, \bibinfo {author} {\bibfnamefont {Q.}~\bibnamefont {Li}}, \bibinfo
  {author} {\bibfnamefont {X.-L.}\ \bibnamefont {Wen}}, \bibinfo {author}
  {\bibfnamefont {J.-Y.}\ \bibnamefont {Zhou}}, \bibinfo {author}
  {\bibfnamefont {W.-X.}\ \bibnamefont {Lin}}, \bibinfo {author} {\bibfnamefont
  {Z.-H.}\ \bibnamefont {Hao}}, \bibinfo {author} {\bibfnamefont {J.-S.}\
  \bibnamefont {Xu}}, \bibinfo {author} {\bibfnamefont {C.-F.}\ \bibnamefont
  {Li}},\ and\ \bibinfo {author} {\bibfnamefont {G.-C.}\ \bibnamefont {Guo}},\
  }\bibfield  {title} {\bibinfo {title} {{Maskless Generation of Single Silicon
  Vacancy Arrays in Silicon Carbide by a Focused He+ Ion Beam}},\ }\href
  {https://pubs.acs.org/doi/abs/10.1021/acsphotonics.2c01209} {\bibfield
  {journal} {\bibinfo  {journal} {ACS Publications}\ }}\Eprint
  {https://arxiv.org/abs/2209.08505v1} {arXiv:2209.08505v1} \BibitemShut
  {NoStop}%
\bibitem [{\citenamefont {Schr{\"o}der}\ \emph {et~al.}(2017)\citenamefont
  {Schr{\"o}der}, \citenamefont {Trusheim}, \citenamefont {Walsh},
  \citenamefont {Li}, \citenamefont {Zheng}, \citenamefont {Schukraft},
  \citenamefont {Sipahigil}, \citenamefont {Evans}, \citenamefont {Sukachev},
  \citenamefont {Nguyen}, \citenamefont {Pacheco}, \citenamefont {Camacho},
  \citenamefont {Bielejec}, \citenamefont {Lukin},\ and\ \citenamefont
  {Englund}}]{schroder2017scalable}%
  \BibitemOpen
  \bibfield  {author} {\bibinfo {author} {\bibfnamefont {T.}~\bibnamefont
  {Schr{\"o}der}}, \bibinfo {author} {\bibfnamefont {M.~E.}\ \bibnamefont
  {Trusheim}}, \bibinfo {author} {\bibfnamefont {M.}~\bibnamefont {Walsh}},
  \bibinfo {author} {\bibfnamefont {L.}~\bibnamefont {Li}}, \bibinfo {author}
  {\bibfnamefont {J.}~\bibnamefont {Zheng}}, \bibinfo {author} {\bibfnamefont
  {M.}~\bibnamefont {Schukraft}}, \bibinfo {author} {\bibfnamefont
  {A.}~\bibnamefont {Sipahigil}}, \bibinfo {author} {\bibfnamefont {R.~E.}\
  \bibnamefont {Evans}}, \bibinfo {author} {\bibfnamefont {D.~D.}\ \bibnamefont
  {Sukachev}}, \bibinfo {author} {\bibfnamefont {C.~T.}\ \bibnamefont
  {Nguyen}}, \bibinfo {author} {\bibfnamefont {J.~L.}\ \bibnamefont {Pacheco}},
  \bibinfo {author} {\bibfnamefont {R.~M.}\ \bibnamefont {Camacho}}, \bibinfo
  {author} {\bibfnamefont {E.~S.}\ \bibnamefont {Bielejec}}, \bibinfo {author}
  {\bibfnamefont {M.~D.}\ \bibnamefont {Lukin}},\ and\ \bibinfo {author}
  {\bibfnamefont {D.}~\bibnamefont {Englund}},\ }\bibfield  {title} {\bibinfo
  {title} {Scalable focused ion beam creation of nearly lifetime-limited single
  quantum emitters in diamond nanostructures},\ }\href@noop {} {\bibfield
  {journal} {\bibinfo  {journal} {Nature communications}\ }\textbf {\bibinfo
  {volume} {8}},\ \bibinfo {pages} {1} (\bibinfo {year} {2017})}\BibitemShut
  {NoStop}%
\bibitem [{\citenamefont {Chen}\ \emph {et~al.}(2017)\citenamefont {Chen},
  \citenamefont {Salter}, \citenamefont {Knauer}, \citenamefont {Weng},
  \citenamefont {Frangeskou}, \citenamefont {Stephen}, \citenamefont {Ishmael},
  \citenamefont {Dolan}, \citenamefont {Johnson}, \citenamefont {Green},
  \citenamefont {Morley}, \citenamefont {Newton}, \citenamefont {Rarity},
  \citenamefont {Booth},\ and\ \citenamefont {Smith}}]{Chen2017}%
  \BibitemOpen
  \bibfield  {author} {\bibinfo {author} {\bibfnamefont {Y.-C.}\ \bibnamefont
  {Chen}}, \bibinfo {author} {\bibfnamefont {P.~S.}\ \bibnamefont {Salter}},
  \bibinfo {author} {\bibfnamefont {S.}~\bibnamefont {Knauer}}, \bibinfo
  {author} {\bibfnamefont {L.}~\bibnamefont {Weng}}, \bibinfo {author}
  {\bibfnamefont {A.~C.}\ \bibnamefont {Frangeskou}}, \bibinfo {author}
  {\bibfnamefont {C.~J.}\ \bibnamefont {Stephen}}, \bibinfo {author}
  {\bibfnamefont {S.~N.}\ \bibnamefont {Ishmael}}, \bibinfo {author}
  {\bibfnamefont {P.~R.}\ \bibnamefont {Dolan}}, \bibinfo {author}
  {\bibfnamefont {S.}~\bibnamefont {Johnson}}, \bibinfo {author} {\bibfnamefont
  {B.~L.}\ \bibnamefont {Green}}, \bibinfo {author} {\bibfnamefont {G.~W.}\
  \bibnamefont {Morley}}, \bibinfo {author} {\bibfnamefont {M.~E.}\
  \bibnamefont {Newton}}, \bibinfo {author} {\bibfnamefont {J.~G.}\
  \bibnamefont {Rarity}}, \bibinfo {author} {\bibfnamefont {M.~J.}\
  \bibnamefont {Booth}},\ and\ \bibinfo {author} {\bibfnamefont {J.~M.}\
  \bibnamefont {Smith}},\ }\bibfield  {title} {\bibinfo {title} {{Laser writing
  of coherent colour centres in diamond}},\ }\bibfield  {journal} {\bibinfo
  {journal} {Nature Photonics}\ }\textbf {\bibinfo {volume} {11}},\ \href
  {https://doi.org/10.1038/NPHOTON.2016.234} {10.1038/NPHOTON.2016.234}
  (\bibinfo {year} {2017})\BibitemShut {NoStop}%
\bibitem [{\citenamefont {Chen}\ \emph {et~al.}(2019)\citenamefont {Chen},
  \citenamefont {Salter}, \citenamefont {Niethammer}, \citenamefont {Widmann},
  \citenamefont {Kaiser}, \citenamefont {Nagy}, \citenamefont {Morioka},
  \citenamefont {Babin}, \citenamefont {Erlekampf}, \citenamefont {Berwian},
  \citenamefont {Booth},\ and\ \citenamefont {Wrachtrup}}]{Chen2019}%
  \BibitemOpen
  \bibfield  {author} {\bibinfo {author} {\bibfnamefont {Y.~C.}\ \bibnamefont
  {Chen}}, \bibinfo {author} {\bibfnamefont {P.~S.}\ \bibnamefont {Salter}},
  \bibinfo {author} {\bibfnamefont {M.}~\bibnamefont {Niethammer}}, \bibinfo
  {author} {\bibfnamefont {M.}~\bibnamefont {Widmann}}, \bibinfo {author}
  {\bibfnamefont {F.}~\bibnamefont {Kaiser}}, \bibinfo {author} {\bibfnamefont
  {R.}~\bibnamefont {Nagy}}, \bibinfo {author} {\bibfnamefont {N.}~\bibnamefont
  {Morioka}}, \bibinfo {author} {\bibfnamefont {C.}~\bibnamefont {Babin}},
  \bibinfo {author} {\bibfnamefont {J.}~\bibnamefont {Erlekampf}}, \bibinfo
  {author} {\bibfnamefont {P.}~\bibnamefont {Berwian}}, \bibinfo {author}
  {\bibfnamefont {M.~J.}\ \bibnamefont {Booth}},\ and\ \bibinfo {author}
  {\bibfnamefont {J.}~\bibnamefont {Wrachtrup}},\ }\bibfield  {title} {\bibinfo
  {title} {{Laser Writing of Scalable Single Color Centers in Silicon
  Carbide}},\ }\href
  {https://doi.org/10.1021/ACS.NANOLETT.8B05070/SUPPL_FILE/NL8B05070_SI_001.PDF}
  {\bibfield  {journal} {\bibinfo  {journal} {Nano Letters}\ }\textbf {\bibinfo
  {volume} {19}},\ \bibinfo {pages} {2377} (\bibinfo {year} {2019})},\ \Eprint
  {https://arxiv.org/abs/1812.04284} {arXiv:1812.04284} \BibitemShut {NoStop}%
\bibitem [{\citenamefont {Almutairi}\ \emph {et~al.}(2022)\citenamefont
  {Almutairi}, \citenamefont {Partridge}, \citenamefont {Xu}, \citenamefont
  {Cole},\ and\ \citenamefont {Holland}}]{Almutairi2022}%
  \BibitemOpen
  \bibfield  {author} {\bibinfo {author} {\bibfnamefont {A.~F.}\ \bibnamefont
  {Almutairi}}, \bibinfo {author} {\bibfnamefont {J.~G.}\ \bibnamefont
  {Partridge}}, \bibinfo {author} {\bibfnamefont {C.}~\bibnamefont {Xu}},
  \bibinfo {author} {\bibfnamefont {I.~S.}\ \bibnamefont {Cole}},\ and\
  \bibinfo {author} {\bibfnamefont {A.~S.}\ \bibnamefont {Holland}},\
  }\bibfield  {title} {\bibinfo {title} {{Direct writing of divacancy centers
  in silicon carbide by femtosecond laser irradiation and subsequent thermal
  annealing}},\ }\href {https://doi.org/10.1063/5.0070014} {\bibfield
  {journal} {\bibinfo  {journal} {Applied Physics Letters}\ }\textbf {\bibinfo
  {volume} {120}},\ \bibinfo {pages} {014003} (\bibinfo {year}
  {2022})}\BibitemShut {NoStop}%
\bibitem [{\citenamefont {Castelletto}\ \emph {et~al.}(2018)\citenamefont
  {Castelletto}, \citenamefont {Almutairi}, \citenamefont {Kumagai},
  \citenamefont {Katkus}, \citenamefont {Hayasaki}, \citenamefont {Johnson},\
  and\ \citenamefont {Juodkazis}}]{Almutairi2018}%
  \BibitemOpen
  \bibfield  {author} {\bibinfo {author} {\bibfnamefont {S.}~\bibnamefont
  {Castelletto}}, \bibinfo {author} {\bibfnamefont {A.~F.~M.}\ \bibnamefont
  {Almutairi}}, \bibinfo {author} {\bibfnamefont {K.}~\bibnamefont {Kumagai}},
  \bibinfo {author} {\bibfnamefont {T.}~\bibnamefont {Katkus}}, \bibinfo
  {author} {\bibfnamefont {Y.}~\bibnamefont {Hayasaki}}, \bibinfo {author}
  {\bibfnamefont {B.~C.}\ \bibnamefont {Johnson}},\ and\ \bibinfo {author}
  {\bibfnamefont {S.}~\bibnamefont {Juodkazis}},\ }\bibfield  {title} {\bibinfo
  {title} {{Photoluminescence in hexagonal silicon carbide by direct
  femtosecond laser writing}},\ }\href {https://doi.org/10.1364/OL.43.006077}
  {\bibfield  {journal} {\bibinfo  {journal} {Optics Letters}\ }\textbf
  {\bibinfo {volume} {43}},\ \bibinfo {pages} {6077} (\bibinfo {year}
  {2018})}\BibitemShut {NoStop}%
\bibitem [{\citenamefont {Dietz}\ \emph {et~al.}()\citenamefont {Dietz},
  \citenamefont {Jiang}, \citenamefont {Day}, \citenamefont {Bhave},\ and\
  \citenamefont {Hu}}]{Dietz2022}%
  \BibitemOpen
  \bibfield  {author} {\bibinfo {author} {\bibfnamefont {J.~R.}\ \bibnamefont
  {Dietz}}, \bibinfo {author} {\bibfnamefont {B.}~\bibnamefont {Jiang}},
  \bibinfo {author} {\bibfnamefont {A.~M.}\ \bibnamefont {Day}}, \bibinfo
  {author} {\bibfnamefont {S.~A.}\ \bibnamefont {Bhave}},\ and\ \bibinfo
  {author} {\bibfnamefont {E.~L.}\ \bibnamefont {Hu}},\ }\bibfield  {title}
  {\bibinfo {title} {{Spin-Acoustic Control of Silicon Vacancies in 4H Silicon
  Carbide}},\ }\href@noop {} {\bibfield  {journal} {\bibinfo  {journal} {arXiv
  2205.15488}\ }}\Eprint {https://arxiv.org/abs/2205.15488v1}
  {arXiv:2205.15488v1} \BibitemShut {NoStop}%
\bibitem [{\citenamefont {Whiteley}\ \emph {et~al.}(2019)\citenamefont
  {Whiteley}, \citenamefont {Wolfowicz}, \citenamefont {Anderson},
  \citenamefont {Bourassa}, \citenamefont {Ma}, \citenamefont {Ye},
  \citenamefont {Koolstra}, \citenamefont {Satzinger}, \citenamefont {Holt},
  \citenamefont {Heremans}, \citenamefont {Cleland}, \citenamefont {Schuster},
  \citenamefont {Galli},\ and\ \citenamefont {Awschalom}}]{Whiteley2019}%
  \BibitemOpen
  \bibfield  {author} {\bibinfo {author} {\bibfnamefont {S.~J.}\ \bibnamefont
  {Whiteley}}, \bibinfo {author} {\bibfnamefont {G.}~\bibnamefont {Wolfowicz}},
  \bibinfo {author} {\bibfnamefont {C.~P.}\ \bibnamefont {Anderson}}, \bibinfo
  {author} {\bibfnamefont {A.}~\bibnamefont {Bourassa}}, \bibinfo {author}
  {\bibfnamefont {H.}~\bibnamefont {Ma}}, \bibinfo {author} {\bibfnamefont
  {M.}~\bibnamefont {Ye}}, \bibinfo {author} {\bibfnamefont {G.}~\bibnamefont
  {Koolstra}}, \bibinfo {author} {\bibfnamefont {K.~J.}\ \bibnamefont
  {Satzinger}}, \bibinfo {author} {\bibfnamefont {M.~V.}\ \bibnamefont {Holt}},
  \bibinfo {author} {\bibfnamefont {F.~J.}\ \bibnamefont {Heremans}}, \bibinfo
  {author} {\bibfnamefont {A.~N.}\ \bibnamefont {Cleland}}, \bibinfo {author}
  {\bibfnamefont {D.~I.}\ \bibnamefont {Schuster}}, \bibinfo {author}
  {\bibfnamefont {G.}~\bibnamefont {Galli}},\ and\ \bibinfo {author}
  {\bibfnamefont {D.~D.}\ \bibnamefont {Awschalom}},\ }\bibfield  {title}
  {\bibinfo {title} {{Spin–phonon interactions in silicon carbide addressed
  by Gaussian acoustics}},\ }\bibfield  {journal} {\bibinfo  {journal} {Nature
  Physics}\ }\textbf {\bibinfo {volume} {15}},\ \href
  {https://doi.org/10.1038/s41567-019-0420-0} {10.1038/s41567-019-0420-0}
  (\bibinfo {year} {2019})\BibitemShut {NoStop}%
\bibitem [{\citenamefont {Anderson}\ \emph {et~al.}(2022)\citenamefont
  {Anderson}, \citenamefont {Glen}, \citenamefont {Zeledon}, \citenamefont
  {Bourassa}, \citenamefont {Jin}, \citenamefont {Zhu}, \citenamefont
  {Vorwerk}, \citenamefont {Crook}, \citenamefont {Abe}, \citenamefont
  {Ul-Hassan}, \citenamefont {Ohshima}, \citenamefont {Son}, \citenamefont
  {Galli},\ and\ \citenamefont {Awschalom}}]{Anderson2022}%
  \BibitemOpen
  \bibfield  {author} {\bibinfo {author} {\bibfnamefont {C.~P.}\ \bibnamefont
  {Anderson}}, \bibinfo {author} {\bibfnamefont {E.~O.}\ \bibnamefont {Glen}},
  \bibinfo {author} {\bibfnamefont {C.}~\bibnamefont {Zeledon}}, \bibinfo
  {author} {\bibfnamefont {A.}~\bibnamefont {Bourassa}}, \bibinfo {author}
  {\bibfnamefont {Y.}~\bibnamefont {Jin}}, \bibinfo {author} {\bibfnamefont
  {Y.}~\bibnamefont {Zhu}}, \bibinfo {author} {\bibfnamefont {C.}~\bibnamefont
  {Vorwerk}}, \bibinfo {author} {\bibfnamefont {A.~L.}\ \bibnamefont {Crook}},
  \bibinfo {author} {\bibfnamefont {H.}~\bibnamefont {Abe}}, \bibinfo {author}
  {\bibfnamefont {J.}~\bibnamefont {Ul-Hassan}}, \bibinfo {author}
  {\bibfnamefont {T.}~\bibnamefont {Ohshima}}, \bibinfo {author} {\bibfnamefont
  {N.~T.}\ \bibnamefont {Son}}, \bibinfo {author} {\bibfnamefont
  {G.}~\bibnamefont {Galli}},\ and\ \bibinfo {author} {\bibfnamefont {D.~D.}\
  \bibnamefont {Awschalom}},\ }\bibfield  {title} {\bibinfo {title}
  {{Five-second coherence of a single spin with single-shot readout in silicon
  carbide}},\ }\href
  {https://doi.org/10.1126/SCIADV.ABM5912/SUPPL_FILE/SCIADV.ABM5912_SM.PDF}
  {\bibfield  {journal} {\bibinfo  {journal} {Science Advances}\ }\textbf
  {\bibinfo {volume} {8}},\ \bibinfo {pages} {5912} (\bibinfo {year} {2022})},\
  \Eprint {https://arxiv.org/abs/2110.01590} {arXiv:2110.01590} \BibitemShut
  {NoStop}%
\bibitem [{\citenamefont {Lukin}\ \emph
  {et~al.}(2020{\natexlab{b}})\citenamefont {Lukin}, \citenamefont {Guidry},\
  and\ \citenamefont {Vu{\v{c}}kovic}}]{Lukin2020}%
  \BibitemOpen
  \bibfield  {author} {\bibinfo {author} {\bibfnamefont {D.~M.}\ \bibnamefont
  {Lukin}}, \bibinfo {author} {\bibfnamefont {M.~A.}\ \bibnamefont {Guidry}},\
  and\ \bibinfo {author} {\bibfnamefont {J.}~\bibnamefont {Vu{\v{c}}kovic}},\
  }\bibfield  {title} {\bibinfo {title} {{Integrated Quantum Photonics with
  Silicon Carbide: Challenges and Prospects}},\ }\href
  {https://doi.org/10.1103/PRXQuantum.1.020102} {\bibfield  {journal} {\bibinfo
   {journal} {Physical Review Applied}\ }\textbf {\bibinfo {volume} {10}},\
  \bibinfo {pages} {20102} (\bibinfo {year} {2020}{\natexlab{b}})}\BibitemShut
  {NoStop}%
\bibitem [{\citenamefont {Dietz}\ and\ \citenamefont {Hu}(2022)}]{Dietz2022a}%
  \BibitemOpen
  \bibfield  {author} {\bibinfo {author} {\bibfnamefont {J.~R.}\ \bibnamefont
  {Dietz}}\ and\ \bibinfo {author} {\bibfnamefont {E.~L.}\ \bibnamefont {Hu}},\
  }\bibfield  {title} {\bibinfo {title} {{Optical and strain stabilization of
  point defects in silicon carbide}},\ }\href
  {https://doi.org/10.1063/5.0087805} {\bibfield  {journal} {\bibinfo
  {journal} {Applied Physics Letters}\ }\textbf {\bibinfo {volume} {120}},\
  \bibinfo {pages} {184001} (\bibinfo {year} {2022})}\BibitemShut {NoStop}%
\bibitem [{\citenamefont {Widmann}\ \emph {et~al.}(2018)\citenamefont
  {Widmann}, \citenamefont {Niethammer}, \citenamefont {Makino}, \citenamefont
  {Rendler}, \citenamefont {Lasse}, \citenamefont {Ohshima}, \citenamefont {{Ul
  Hassan}}, \citenamefont {{Tien Son}}, \citenamefont {Lee},\ and\
  \citenamefont {Wrachtrup}}]{Widmann2018}%
  \BibitemOpen
  \bibfield  {author} {\bibinfo {author} {\bibfnamefont {M.}~\bibnamefont
  {Widmann}}, \bibinfo {author} {\bibfnamefont {M.}~\bibnamefont {Niethammer}},
  \bibinfo {author} {\bibfnamefont {T.}~\bibnamefont {Makino}}, \bibinfo
  {author} {\bibfnamefont {T.}~\bibnamefont {Rendler}}, \bibinfo {author}
  {\bibfnamefont {S.}~\bibnamefont {Lasse}}, \bibinfo {author} {\bibfnamefont
  {T.}~\bibnamefont {Ohshima}}, \bibinfo {author} {\bibfnamefont
  {J.}~\bibnamefont {{Ul Hassan}}}, \bibinfo {author} {\bibfnamefont
  {N.}~\bibnamefont {{Tien Son}}}, \bibinfo {author} {\bibfnamefont {S.-Y.}\
  \bibnamefont {Lee}},\ and\ \bibinfo {author} {\bibfnamefont {J.}~\bibnamefont
  {Wrachtrup}},\ }\bibfield  {title} {\bibinfo {title} {{Bright single photon
  sources in lateral silicon carbide light emitting diodes}},\ }\bibfield
  {journal} {\bibinfo  {journal} {Applied Physics Letters}\ }\textbf {\bibinfo
  {volume} {112}},\ \href {https://doi.org/10.1063/1.5032291}
  {10.1063/1.5032291} (\bibinfo {year} {2018})\BibitemShut {NoStop}%
\bibitem [{Day()}]{Day}%
  \BibitemOpen
  \href@noop {} {\bibinfo {title} {Supplementary information}}\BibitemShut
  {NoStop}%
\bibitem [{\citenamefont {{Kuate Defo}}\ \emph {et~al.}(2018)\citenamefont
  {{Kuate Defo}}, \citenamefont {Zhang}, \citenamefont {Bracher}, \citenamefont
  {Kim}, \citenamefont {Hu},\ and\ \citenamefont {Kaxiras}}]{KuateDefo2018}%
  \BibitemOpen
  \bibfield  {author} {\bibinfo {author} {\bibfnamefont {R.}~\bibnamefont
  {{Kuate Defo}}}, \bibinfo {author} {\bibfnamefont {X.}~\bibnamefont {Zhang}},
  \bibinfo {author} {\bibfnamefont {D.}~\bibnamefont {Bracher}}, \bibinfo
  {author} {\bibfnamefont {G.}~\bibnamefont {Kim}}, \bibinfo {author}
  {\bibfnamefont {E.}~\bibnamefont {Hu}},\ and\ \bibinfo {author}
  {\bibfnamefont {E.}~\bibnamefont {Kaxiras}},\ }\bibfield  {title} {\bibinfo
  {title} {{Energetics and kinetics of vacancy defects in 4H -SiC}},\
  }\bibfield  {journal} {\bibinfo  {journal} {Physical Review B}\ }\textbf
  {\bibinfo {volume} {98}},\ \href {https://doi.org/10.1103/PHYSREVB.98.104103}
  {10.1103/PHYSREVB.98.104103} (\bibinfo {year} {2018})\BibitemShut {NoStop}%
\bibitem [{\citenamefont {Castelletto}\ \emph {et~al.}(2014)\citenamefont
  {Castelletto}, \citenamefont {Johnson}, \citenamefont {Iv{\'{a}}dy},
  \citenamefont {Stavrias}, \citenamefont {Umeda}, \citenamefont {Gali},\ and\
  \citenamefont {Ohshima}}]{Castelletto2014}%
  \BibitemOpen
  \bibfield  {author} {\bibinfo {author} {\bibfnamefont {S.}~\bibnamefont
  {Castelletto}}, \bibinfo {author} {\bibfnamefont {B.~C.}\ \bibnamefont
  {Johnson}}, \bibinfo {author} {\bibfnamefont {V.}~\bibnamefont
  {Iv{\'{a}}dy}}, \bibinfo {author} {\bibfnamefont {N.}~\bibnamefont
  {Stavrias}}, \bibinfo {author} {\bibfnamefont {T.}~\bibnamefont {Umeda}},
  \bibinfo {author} {\bibfnamefont {A.}~\bibnamefont {Gali}},\ and\ \bibinfo
  {author} {\bibfnamefont {T.}~\bibnamefont {Ohshima}},\ }\bibfield  {title}
  {\bibinfo {title} {{A silicon carbide room-temperature single-photon
  source}},\ }\href {https://doi.org/10.1038/nmat3806} {\bibfield  {journal}
  {\bibinfo  {journal} {Nature Materials}\ }\textbf {\bibinfo {volume} {13}},\
  \bibinfo {pages} {151} (\bibinfo {year} {2014})}\BibitemShut {NoStop}%
\bibitem [{\citenamefont {Gattass}\ and\ \citenamefont
  {Mazur}(2008)}]{Gattass2008}%
  \BibitemOpen
  \bibfield  {author} {\bibinfo {author} {\bibfnamefont {R.~R.}\ \bibnamefont
  {Gattass}}\ and\ \bibinfo {author} {\bibfnamefont {E.}~\bibnamefont
  {Mazur}},\ }\bibfield  {title} {\bibinfo {title} {{Femtosecond laser
  micromachining in transparent materials}},\ }\href
  {https://www.nature.com/articles/nphoton.2008.47} {\bibfield  {journal}
  {\bibinfo  {journal} {Nature Photonics}\ }\textbf {\bibinfo {volume} {2}},\
  \bibinfo {pages} {219} (\bibinfo {year} {2008})}\BibitemShut {NoStop}%
\bibitem [{\citenamefont {Skowronski}\ \emph {et~al.}(2002)\citenamefont
  {Skowronski}, \citenamefont {Liu}, \citenamefont {Vetter}, \citenamefont
  {Dudley}, \citenamefont {Hallin},\ and\ \citenamefont
  {Lendenmann}}]{Skowronski2002}%
  \BibitemOpen
  \bibfield  {author} {\bibinfo {author} {\bibfnamefont {M.}~\bibnamefont
  {Skowronski}}, \bibinfo {author} {\bibfnamefont {J.~Q.}\ \bibnamefont {Liu}},
  \bibinfo {author} {\bibfnamefont {W.~M.}\ \bibnamefont {Vetter}}, \bibinfo
  {author} {\bibfnamefont {M.}~\bibnamefont {Dudley}}, \bibinfo {author}
  {\bibfnamefont {C.}~\bibnamefont {Hallin}},\ and\ \bibinfo {author}
  {\bibfnamefont {H.}~\bibnamefont {Lendenmann}},\ }\bibfield  {title}
  {\bibinfo {title} {Recombination-enhanced defect motion in forward-biased
  4h-sic p-n diodes},\ }\href@noop {} {\bibfield  {journal} {\bibinfo
  {journal} {Journal of applied physics}\ }\textbf {\bibinfo {volume} {92}},\
  \bibinfo {pages} {4699} (\bibinfo {year} {2002})}\BibitemShut {NoStop}%
\bibitem [{\citenamefont {Kimerling}(1978)}]{Kimerling1978}%
  \BibitemOpen
  \bibfield  {author} {\bibinfo {author} {\bibfnamefont {L.~C.}\ \bibnamefont
  {Kimerling}},\ }\bibfield  {title} {\bibinfo {title} {{Recombination enhanced
  defect reactions}},\ }\bibfield  {journal} {\bibinfo  {journal} {Solid State
  Electronics}\ }\textbf {\bibinfo {volume} {21}},\ \href
  {https://doi.org/10.1016/0038-1101(78)90215-0} {10.1016/0038-1101(78)90215-0}
  (\bibinfo {year} {1978})\BibitemShut {NoStop}%
\bibitem [{\citenamefont {Lang}\ and\ \citenamefont
  {Kimerling}(1974)}]{Lang1974}%
  \BibitemOpen
  \bibfield  {author} {\bibinfo {author} {\bibfnamefont {D.~V.}\ \bibnamefont
  {Lang}}\ and\ \bibinfo {author} {\bibfnamefont {L.~C.}\ \bibnamefont
  {Kimerling}},\ }\bibfield  {title} {\bibinfo {title} {Observation of
  recombination-enhanced defect reactions in semiconductors},\ }\href@noop {}
  {\bibfield  {journal} {\bibinfo  {journal} {Physical Review Letters}\
  }\textbf {\bibinfo {volume} {33}},\ \bibinfo {pages} {489} (\bibinfo {year}
  {1974})}\BibitemShut {NoStop}%
\end{thebibliography}%


%

\end{document}